\documentclass[12pt]{article}
\renewcommand{\baselinestretch}{1.5} 

\newcommand{\bm}[1]{\mbox{\boldmath $#1$}}

\newcommand{\epsi}{\bm{\varepsilon_i}}
\newcommand{\bepsi}{\bm{\varepsilon}}

\newcommand{\Cov}{\mbox{Cov}}
\newcommand{\bitem}{\begin{itemize}}
\newcommand{\eitem}{\end{itemize}}

 \newcommand{\beps}{\mbox{\boldmath $\varepsilon$}}

 \newcommand{\bSigma}{\mbox{\boldmath $\Sigma$}}

 \newcommand{\BA}{\mbox{\boldmath $A$}}

 \newcommand{\BR}{\mbox{\boldmath $R$}}

 \newcommand{\BT}{\mbox{\boldmath $T$}}

 \newcommand{\BX}{\mbox{\boldmath $X$}}

 \newcommand{\BJ}{\mbox{\boldmath $J$}}
 \newcommand{\BI}{\mbox{\boldmath $I$}}
 \newcommand{\bx}{\mbox{\boldmath $x$}}
 \newcommand{\bp}{\mbox{\boldmath $p$}}

 \newcommand{\bfSigma}{\mbox{\boldmath $\Sigma$}}

 \newcommand{\bfvarepsilon}{\mbox{\boldmath $\varepsilon$}}

\begin{document}

\centerline{\large\bf Estimating Clonality}

\begin{center} 
{\bf Yi Liu}$^{1,3}$, {\bf Andrew Z. Fire}$^{2,3}$, {\bf Scott Boyd}$^3$ and {\bf Richard A. Olshen}$^{4^*}$ \\
$^1$Bioinformatics Training Program, Stanford University; \\ 
$^2$Department of Genetics, Stanford University School of Medicine \\
$^3$Department of Pathology, Stanford University School of Medicine \\
$^{4^*}$Division of Biostatistics, Stanford University School of Medicine \\
{\it email$^*$}: olshen@stanford.edu
\end{center}

SUMMARY: Challenges of assessing complexity and clonality in populations of mixed species arise in diverse areas of modern biology, including estimating diversity and clonality in microbiome populations, measuring patterns of T and B cell clonality, and determining the underlying tumor cell population structure in cancer.  Here we address the problem of quantifying populations, with our analysis directed toward systems for which previously defined algorithms allow the sequence-based identification of clonal subpopulations.  Data come from replicate sequencing libraries generated from a sample, potentially with very different depths.  While certain properties of the underlying clonal distribution (most notably the total number of clones) are difficult to estimate accurately from data representing a small fraction of the total population, the population-level ``clonality'' metric that is the sum of squared probabilities of the respective species can be calculated.  (This is the sum of squared entries of a high-dimensional vector $\bp$ of relative frequencies.)  The clonality score is the probability of a clonal relationship between two randomly chosen members of the population of interest.  A principal takeaway message is that knowing a functional of $\bp$ well may not depend on knowing $\bp$ itself very well.  

Our work has led to software, which we call {\it lymphclon}; it has been deposited in the CRAN library.


\section{Introduction}

Understanding the population structures of biological entities is an important goal in many areas of medical research.  The goals of defining the repertoires of lymphocytes that mediate healthy immune responses and immunological diseases are many.  They include characterizing the microbial populations inhabiting various tissue sites in the human body in normal and aberrant physiology and detecting the sub-populations of genetically distinct tumor cells in a cancer that may respond differently to treatment.  All efforts have benefited from recent advances in DNA sequencing technology.  This paper is our contribution to estimating a particular functional, {\it clonality}, of the relative frequencies of species within any mixed biological population in which common genetic or clonal origins can be determined pairwise for individuals in the population.  The metric that we study is the Gini-Simpson index of the vector of those relative frequencies (see, for example, Breiman et al., 1984), which amounts to the probability that two independently chosen rearrangements belong to the same clone.  Our methodology involves empirical Bayes estimation (see Efron, 2010, and references therein), and jackknifing (see Miller, 1974) at two key points.  Although the jackknifing we do does not yield unbiased estimators of clonality, the estimators are close enough to being unbiased that we can employ a formula we give for best linear unbiased estimation with arbitrary positive definite covariance.  The formula involves the inverse of that positive definite matrix.  Regularization implicit in our jackknifing entails something counterintuitive, namely that simultaneously we can estimate the set of eigenvalues of certain covariances and their inverses well enough.  In what follows, we include the model and estimation of {\it clonality} that devolves from it followed by some results of simulations.  In these simulations we found that the mean squared error of estimated clonality can be reduced by more than 98\% that of another recent estimator, perhaps the best currently available in immunology (Parawesmaran et al., 2013).  A computer package for implementation of our ideas in termed {\it lymphclon}.  It has been deposited in the CRAN library.

Because {\it clonality} is a quadratic function of the cited relative frequencies, its mean squared error involves fourth moments of those frequencies (see also two recent papers by Jiao, Venkat, and Weissman (2014)). These fourth moments are computed for a model that enables estimation.  The model has tricky consequences; one very special case of the problem we cite is that of estimating the squared length of a Euclidean multinomial vector from its frequencies.

Our problem has arisen in many contexts, including but not limited to ecology.  The reader is encouraged to read the works of Chao (see her 1989 paper and its references), that deal with estimating what amounts to the vector of relative frequencies $\bp$ itself.  Our work shows that one can estimate such functionals of $\bp$ as {\it clonality} well without knowing $\bp$ itself very well.  

To provide a concrete example, we choose one application for the purposes of discussion and for refining estimates of clonality.  B cells of the adaptive immune system are large populations of cells with characteristic and highly diverse genomic rearrangements of gene segments at loci encoding antigen receptor proteins, called antibodies (or immunoglobulins).  In response to antigens associated with pathogens and other environmental exposures, B cells with receptors recognizing the antigens undergo extensive clonal expansions, giving rise to many progeny.  We assume, albeit implicitly, that there is a correspondence between DNA sequence incidence for individual clonal subpopulations in an underlying population relative frequencies with which the corresponding and characteristic DNA sequences are represented in the pool of a series of replicate amplifications.  These {\it replicates} are subsamples of a sample of blood taken from an individual.  The antigen receptor rearrangements are amplified by the polymerase chain reaction (PCR) and sequenced.

\section{The Model, Assumptions and Implications}

\subsection{Basic assumptions \label{basicAssumptions}}
We fix a cell type, say, B cells, which can each possess one of many different antibody gene rearrangements.  We focus on an individual person from whom B cells are sampled, with his or her random probability vector being designated by $\bp$.  
Write $\bp = (\bp(1), \bp (2), ...)$.  Because we study functionals of $\bp$ that are invariant to permutations its coordinates, without loss we could assume that $\bp (1) \ge \bp (2) \ge ...$  Readers will see in a later section that motivation for an empirical Bayes argument we employ makes this string of inequalities implausible.  For all $j= 1, 2, ..., \bp (j) \ge 0$.  The cardinality of $\{ j: \bp(j) > 0\}$ is denoted by $C$.  For technical reasons that do not merit discussion here, we require that $C \ge 3$.  In practice $C \ge 10^6$.  $C$ denotes {\it richness}, but that parameter is not studied explicitly here.  In any case, $\bp \in {\cal R}^C$; that is, $\bp$ belongs to the Euclidean space ${\cal R}^C$. Write $\bp_i$ for the i$^{th}$ replicate; $i = 1, 2, ..., n$.  We assume that $E(\bp_i | \bp , C)=\bp$. In an obvious notation, $\{ j: \bp_i (j) > 0\}$ is defined to be $C_i$.  Always $C_i < C$; in practice, $C_i \ll C$. 

It is helpful to employ inner product notation.  Thus, if ${\bf 1} \in {\cal R}^C$, and $1(j) = 1$ for $j= 1, ... C$, then $(\bp, {\bf 1}) = 1$.  {\it Clonality} is $(\bp, \bp)$, which we define to be $\theta; \theta = || \bp ||^2_2$.

Write $\bp_i = \bp + \bepsi_i$.  The $C$-dimensional vector $\epsi$ is the ``error'' in the $i^{th}$ replicate.  In biological parlance, the number of {\it reads} in $\bp_i$ is $C_i \Sigma^C_{k=1} \bp_i (k)$.  We will place conditional and unconditional moment constraints on $\{\epsi \}$, but they are not sufficient to determine exact distributions of $\{\epsi \}$.  If they were, then in view of sufficiency, without loss we could take $n=1$.

Because $E(\bp_i) = \bp$, one obvious estimate of $\theta$ is $n^{-1}\ \Sigma_{i=1}^n\ || \bp_i ||^2_2$.  However, the real-valued function $x \rightarrow x^2$ is convex.  So, Jensen's inequality implies that the cited average is a biased, typically extremely biased, estimator of $\theta$.

Were the coordinates of $\bp$ independent, $N(\theta_i, 1)$, say, then $\theta$ would be a non-centrality parameter.  We know, for example from (Perlman and Rasmussen, 1975), how difficult estimating $\theta$ would be in that case.  Because $\bp$ takes values in a unit simplex, a compact subset of ${\cal R}^C$, one might think that the problem we study here is easier than that of estimating a non-centrality parameter.  Perhaps unfortunately, this seems not to be the case.

\subsection{Assumptions regarding errors}

Assumptions on $\{ \beps_i\}$ are tricky, but they are what allow inference on $\theta$.  It pays to recall the definition of $\{ \beps \}$.  A sample of blood is taken from an individual.  It is divided into disjoint subsets.  Separately, they are implied by the polymerase chain reaction (PCR).  Subsequently, what has been amplified is sequenced by the latest technology.  It is assumed that there is a 1-1 correspondence between sequences and rearrangements.  In particular, we assume that
\begin{equation}
E(\beps_j (l) | \bp, \{C_i\}) = 0\ {\rm for\ all}\ (j, l) ; \label{eq1}
\end{equation}
\begin{equation}
P(\varepsilon_j (k) = 0 | \bp (j) = 0) = 1 \ {\rm for\ all}\ (j, k); \label{eq2}
\end{equation}
\begin{equation}
\{ \beps_j\}\ {\rm are\ conditionally\ independent\ given\ } \bp\ \mbox{and}\ \{C_i\}; 
 \mbox{also just given}\ \bp. \label{eq3}
\end{equation}
\begin{equation}
\mbox{for distinct}\ (j, k, l), \beps_j \ {\rm and}\ \beps_k \ \mbox{are conditionally independent given}\
\{\bp, \{C_i\}, \beps_l \} \label{eq4}
\end{equation}
Note that (\ref{eq1}) and $\bp (1) > 0$ imply that, for all $i, \beps_i (1)$ and $\bp(1)$ are negatively correlated.  This is different from assumptions that apply in the usual empirical Bayes argument for the James-Stein estimator of a Guassian mean (cf. Chapter 1 of Efron, 2010).  Our assumption (\ref{eq2}) implies that if a rearrangement is not in the circulation that defines $\bp$, then it is measured without error.  Thus, its relative frequency is $0$.  This assumption may not apply to some technologies or some experiments.  However, we believe that it is the job of both the experimentalist and any first stage of analysis to minimize ``fake'' clones derived from sequencing errors. Further, we believe that this assumption does not invalidate any conclusions drawn from analyses of our data.

For distinct replicates $l$ and $m$, we define
\begin{equation}
\theta_{(l,m)} = \sum^c_{j=1} \ \bp_l (j)\ \bp_m (j) \label{eq5}
\end{equation}
One sees that $E (\theta_{(l,m)} | \bp ) = \theta$.  Likewise, if
\begin{equation}
{\hat\theta}^* = \sum_{l \neq m} \ C_l C_m \theta_{(l,m)} / \sum_{l\neq m} C_l C_m , \label{eq6}
\end{equation}
then $E({\hat\theta}^* | \bp , \{C_i\}) = \theta$.  See (Parameswaran et al., 2013).  Extending ${\hat\theta}^*$ to include summation over the diagonal terms $l = m$ does introduce some bias in estimating $\theta$.  One computes that
\begin{equation}
E( (\bp_i (j))^2 | \bp , \{C_i\}) = (\bp (j))^2 + {\rm Var}\ (\bp_i (j) | \bp, \{C_i \}) \label{eq7}
\end{equation}
But typically $\sum_j {\rm Var}\ (\bp_i (j) | \bp_j, \{C_i \})$ is small.  Thus, bias is not the problem in extending the summation in (\ref{eq6}) to all $(l, m)$ that concerns us.  Instead, it is the (conditional) variance of the sum.  Estimators for which (conditional) bias is little problem but (conditional) variance is reduced are enabled by studying {\it pairs} of $\{ \theta_{(k,l)}\}$.  But, before studying the covariances of such pairs, we require another comment and another assumption.

The definition of $\beps_i$ and simple algebra entail that 
$|| \beps_i ||^2 \le || \bp ||^2_2\ +\ 2 || \bp ||_2 \ ||\ \bp_i ||_2 \ + \ ||\ \bp_i ||^2_2$.  Because
$0 < || \bp ||^2_2 = \sum^C_{j=1} \ (\bp (j))^2 \le \sum^C_{j=1}\ \bp (j) = 1$, and likewise
$|| \bp_i ||^2_2 \le 1$, it follows that 
$E ( || \beps_i ||^2_2 \ |\ \bp, \{C_i\} \le 4$.
Indeed, $\beps$ has positive $l_p$ moments of all orders.  Our assumption is
\begin{eqnarray}
&\mbox {the conditional distribution of}\ ||\beps_i ||^2 \cos^2 (\psi_i) \nonumber \\
&{\rm given}\ \bp \ {\rm and}\ \{C_i\}\ \mbox {does not depend on}\ i . \label{eq8}
\end{eqnarray}
Here $\psi_i$ is the cosine of the angle between $\beps_i$ and $\bp$.

\section{Covariances of Pairs of $\{ \theta_{(k,l)}\}$\label{covariancesPairs}}
It is easy to see that if $(k,l)$ and $(k', l')$ are pairs of indices for which all four subscripts are distinct, then 
$\Cov ( (\theta_{k,l}, \theta_{k', l'}) | \bp, \{C_i\} ) = 0$.  Our next step is to study the covariance in case three indices are distinct, but the fourth is common.  To that end, we compute \\
\begin{eqnarray}
\Cov ( (\theta_{k,l}, \theta_{k', l'})\ |\ \bp, \{C_i\} ) 
 &= & E(\theta_{(k,l)} \theta_{(l,l')}\ |\ \bp , \{ C_i\}) - 
      E(\theta_{(k,l)}\ |\ \bp, \{C_i\})\ E(\theta_{(l,l')}\  |\ \bp, \{C_i\})  \nonumber \\
 &= &E(\theta_{(k,l)} \theta_{(l,l')}\ |\ \bp , \{ C_i\}) - (|| \bp ||^2_2)^2 \nonumber \\
 &= &E(\bp + \beps_k, \bp + \beps_l) (\bp + \beps_l, p + \varepsilon_{l'})\ |\ \bp, \{C_i\}) - ||\bp||^4_2 \nonumber \\ 
 &= &I - II  \label{eq9}
\end{eqnarray}

$I$ is the expectation of the product of two inner products, each inner product having four terms.  Thus, there are 16 terms to $I$.  Simple calculations imply that only two of the 16 terms is not $0$.  In particular,
\begin{eqnarray}
I - II 
 &= &|| \bp ||^4_2 + ||\bp ||^2_2\ E(\cos^2 (\psi_l)\ || \beps_l ||^2_2\ |\ \bp, \{ C_i\} ) - || \bp ||^4_2 \nonumber \\
 &= &|| \bp ||^2_2 \ E(\cos^2 (\psi_l)\ || \beps_l ||^2_2\ |\ \bp, \{ C_i\} ) . \label{eq10}
\end{eqnarray} 

For each $l = 1, 2, ... n$ there are $({{n - 1}\atop 2})$ choices of distinct indices $(k, l')$.  Because we are interested in sums over all sets of pairs $(k, l)$ and $(l, l')$ as described, we are interested in
\begin{eqnarray}
& &({{n-1}\atop 2})\ || \bp ||^2_2 \ \sum^n_{l=1} \ E(\cos^2 (\psi_l) \ ||\beps_l ||^2_2 \ | \ \bp, \{C_i\})  \nonumber \\
&= &[n ({{n-1}\atop 2})\ || \bp ||^2_2]\ [n^{-1}\ \sum^n_{l=1} \ E(\cos^2 (\psi_l) \ ||\beps_l ||^2_2 \ | \ \bp, \{C_i\})]
\label{eq11}
\end{eqnarray}

In view of (\ref{eq3}) and (\ref{eq8}) and comments just prior to (\ref{eq8}), the random variables that appear in the second product in (\ref{eq11}) are law of large numbers averages of independent, identically distributed random variables to which large deviation results (display (4.16) of Hoeffding, 1963 and Section 2 of Varadhan, 2008) apply.  Exact details depend on calculation of Kullback-Leibler information for the distribution in (\ref{eq8}).  However, whatever the exact calculation, there are positive numbers A and B for which, for any $\delta > 0$
\begin{eqnarray}
& &P( | n^{-1} \sum^n_{l=1} \ E(\cos^2 \ (\psi_l)\ ||\ \beps_l ||^2_2 \ | \ \bp, \{C_i\})  \nonumber \\
&- &E (E (\cos^2 (\psi_1)\ || \beps_1 ||^2_2\ |\ \bp, \{ C_i\} | > \delta) < A \exp\{ -Bn \delta \}. 
\label{eq12}
\end{eqnarray}

The term subtracted in (\ref{eq12}) is constant.  The first term in (\ref{eq11}) is irrelevant in view of the last term of (\ref{eq12}). It follows that there are positive constants 
$A'$ and $B'$ for which the product (\ref{eq11}) is within $\delta$ of $E (E (\cos^2 (\psi_1) || \beps_1 ||^2_2\ |\ \bp, \{ C_i\})$ with probability at least $1- \delta$.  Denote the latter conditional expectation by ${\bar\psi}$.

Because there are $n$ replicates, in view of Bonferroni's inequality, there are positive constants $A^{''}$ and $B^{''}$ for which for any $\delta > 0$ and $n$ large enough, the probability that for some replicate the probability (\ref{eq11}) is larger than $\delta$ from ${\bar\psi}$ is less than  $A^{''}$ exp$\{- n B^{''} \delta \}$.  Because a constant is independent of any random variable, in (\ref{eq10}) and (\ref{eq11}) we can replace $\cos^2 (\psi_l)$ by ${\bar\psi}$ (provided we sum over replicates in (\ref{eq10}) and (\ref{eq11}) and thus but for a set of exponentially small probability we can move the constant ${\bar\psi}$ outside the expectation.
 
For our last consideration of this section we study the case $(k, l) = (k', l')$.  Thus, we study 
\begin{eqnarray}
& &Var (\theta_{(k,l)} | \bp, \{C_i \}) = \nonumber \\
& &E(( \bp + \beps_k, \bp + \beps_l ) \ (\bp + \beps_k, \bp + \beps_l ) \ | \ \bp, \{ C_i\}) - 
    E^2 (( \bp + \beps_k, \bp + \beps_l ) \ | \ \bp, \{ C_i\}) . \nonumber
\end{eqnarray}
Obviously, there are ${n \choose 2}$ such terms.  Considerations like those we applied to (\ref{eq10}), (\ref{eq11}), (\ref{eq12}), and the discussion following (\ref{eq12}) entail that there are positive $A^{''}$ and $B^{''}$ for which for any $\delta > 0$ and $n$ big enough, the probability is less than $A^{''} \exp \{ -n B^{''} \delta \}$ that for any pair $(k, l)$, Var$(\theta_{(k,l)} | \bp, \{C_i\})$ departs from $|| \bp ||^2_2\ \psi\ E( || \beps_k ||^2_2 + || \beps_l ||^2_2 \ | \ \bp, \{ C_i\})$ more than $\delta$.

\section{Empirical Bayes Estimation and Jackknifing.}
According to computations in the last section, for pairs of replicates $(k,l), (k', l')$, given $\bp$ and $\{C_i\}$ the conditional covariance of pairs of replicates is 
\begin{equation}
\begin{array}{lll}
0 &= &\mbox{if}\ k, l, k', l' \ \mbox{distinct} \\
|| \bp ||^2_2\ \bar{\psi}\ E(|| \beps_k||^2_2 \ | \ \bp, \{C_i \}) &= &\mbox{if}\ (k,l) \cap (k', l') = \{k\} \\
|| \bp ||^2_2\ \bar{\psi}\ E(|| \beps_k||^2_2 + || \beps_l ||^2_2 \ | \ \bp, \{C_i \}) &= &\mbox{if}\ (k, l) = (k', l') \\ \label{eq13}
\end{array}
\end{equation}
with probability at least $1 - A^{''} \exp \{ -n B^{''} \delta \}$, for any $\delta > 0$ and integer $n$ large enough and fixed constants $A^{''}$ and $B^{''}$.  In the spirit of ordinary empirical Bayes (see Robbins, 1956 and Efron, 2010) we could set an ${n \choose 2}$ by ${n \choose 2}$ covariance matrix ${\bfSigma}$ of pairs of replicates equal to expressions (\ref{eq13}) and solve 
these over-determined, non-linear equations for the $n+2$ constants: 
$|| \bp ||^2_2, {\bar\psi},$ and $E(|| \beps_k ||^2_2 \ | \ \bp, \{C_i \}): k = 1, ... n$.\ 

Obviously, one of the terms in each product of (\ref{eq13}) is $\theta$ itself.  It is part of the multiplier, $\theta {\bar\psi}$.  The path we take has many virtues.  For one, it allows us to ignore these multipliers.  For another, it leads to estimators of $\theta$ that close to being unbiased (given $\bp$ and $\{C_i \}$).  For yet another, these estimators have far smaller variance than does (\ref{eq6}) (see Parameswaron et al., 2013).  This path involves jackknifing estimates of $\theta$ that devolve from estimating a covariance matrix of fixed coordinates of the $n$ replicates.  Obviously, given $\bp$ there are $C$ coordinates in all.  We employ loose empirical Bayes arguments for estimating each of the $n$ terms in the jackknifing.  In passing, we are able to estimate 
$E (||  \bfvarepsilon_k ||^2_2 \ | \ \bp, \{ C_i \}), k = 1, ..., n.$

To begin, suppose that we had $m$ estimators of $\theta \ E_1, ..., E_m$, each nearly unbiased.  Further, suppose that we had an estimator $\tilde{\theta}$ of $\theta$ based on all the data.  Suppose further that
$E (\tilde{\theta} \ | \ \bp, \{ C_i \}) = \theta + (1/n) a_1 + (1/n)^2 a_2 + ...$ 
for some constants $a_1, a_2, ...$  Then we might combine $E_1, ..., E_m$ in a best linear ``unbiased'' way by computing 
\begin{equation}
\hat{\theta} = ({\bx}, \BR {\bf 1}) / || \BR^{-1} {\bf 1} ||^2_2 .
\end{equation}

Here $\BR$ is the $ m \times m $ covariance matrix of $E_1, ... E_m$; ${\bf 1}$ is the m-vector with all entries 1, $\bx$ is the vector with $j^{th}$ entry $E_j, j = 1, ..., m$.  Note that (14) is invariant to constant multiples of $\BR$.  Thus, we can apply (14) if we know $\BR$ only up to constant multiple.  This is an important observation because it allows us to remove the multiplier $\theta \bar{\gamma}$ in (13).

We could eliminate the $(1/n)$ term in the bias of $\tilde{\theta}$ by computing a jackknife estimator (see Miller, 1974).  It is toward that jackknife goal that for empirical Bayes purposes we construct a $C \times n$ matrix $\BX$.  The $n$ columns of $\BX$ are $\bp_1, ..., \bp_n$.  These are, obviously, observed quantities.

Write $\BX_i$ for the $i^{th}$ row of $\BX$, and $\BX_{-i}$ to denote $\{ \BX_l : l \not= i\}$.  (In an ecological context, the $j^{th}$ entry of $\BX_i$ is the measured number of the specie indexed by $i$ in the $j^{th}$ study.)  Because jackknifing matters here, we need to know the covariance matrix, 
$Var (\BX_i | \BX_{-i}, \bp, \{ C_l \}).$  The conditional variance formula (that is, Pythagorean considerations) entail that we write the latter conditional variance as
$$Var (\BX_i | \BX_{-i}, \bp, \{ C_l \}) + Var (E (\BX_i \ | \ \bp, \{C_l \} | \BX_{-i} ).$$
Straightforward algebra shows that the sum can be written 
$Var (\BX_i | \bp, \{ C_l \}) (1 + L)$, where
$L = Var(\bp | \BX_{-i}, \{C_l\}) / \ E(Var (\BX_i | \bp, \{ C_l \})$. 
We could argue -- details omitted here -- that with probability 1 minus a quantity exponentially small, $L$ is $0(1/n)$, with uniformly bounded constant.

Up to the constant of proportionality $\theta \bar{\psi}$, (13) would apply to each of the $n$ terms that figure in jackknifing, if only we had estimators of 
$E (||  \bfvarepsilon_k ||^2_2 \ | \ \bp, \{ C_i \}), k = 1, ..., n$.  Not only are these available in view of (1), but also, we can estimate each term in the ${n\choose 2}  \times {n\choose 2}$ matrix entailed by (13), even with the norming constants.

Rather than estimate the conditional covariance matrix of 
$\BX_i$ given $\BX_{-i}, \bp, \{C_l\}$ directly, we begin the next phase of our arguments by assuming that 
$Var (\BX_i | \BX_{-i}, \bp, \{ C_l \})=$ \\
$E(Var(\BX_i \ | \ \bp, \{C_l \})$, even though the equality is only approximate.  It follows that the conditional covariance matrix can be estimated directly from our data.  In view of (1), a simplified (14), and (4), we can estimate $E (||  \bfvarepsilon_k ||^2 \ | \ \bp, \{ C_l \})$ for every $k$.

We can compute (5) and therefore (6) from data.  Formulas (13) enable us to estimate 
$Var (\hat{\theta}^*)$ directly.  Though we have no mathematical proof, plausibility arguments imply that if we improve estimation by improving the condition number of any estimator of $Var (\BX_i | \BX_{-i}, \bp, \{ C_l \})$, then we should improve the subsequent estimation of $\theta$.

Remember that our basic model is $\bp_i = \bp + \bfvarepsilon_i$; each of $\{ \bp_i\}$, all that we observe, estimates $\bp$, albeit poorly.  We can imagine estimating $\bp$ by its least squares estimator computed for predictors $\{ \bp_i: i=1, ... n\}$.  That least squares estimator has (conditional) covariance matrix $(\BX' \BX)^{-1}$, where the superscripted prime denotes transpose.  If, rather implausibly, the probability distribution of $\{ \beps_l: l = 1, ... n\}$ is exchangeable, that is, the joint distribution is invariant under permutations, then that exchangeability is inherited by each coordinate.  This exchangeability would imply that the $\{ C_l C_{l'} \}$ weighted off diagonal elements of $\BX' \BX$ (as in (6)) should be about the same.  It would follow that $\BX' \BX$ would be approximately $\hat{\theta}^* \BJ$, where $\BJ$ is the $n \times n$ matrix with every entry 1.  This equality would hold for every $i$.  The resulting estimator, common to $i$, might for conventional arguments of unbiasedness (that barely matter here) be $(\hat{\theta}^* - C^{-2}) \BJ$.

Because $\BJ$ has rank only 1 for every {\it n}, and typically $n \gg 1$, we are motivated to improve $\BJ's$ disastrous condition number.  All three approaches we have tried are available with the $R$ command ``corpcor.''  (See Sh\"{a}fer and Strimmer, 2005 and Opgen-Rhein and Strimmer, 2007.)  One regularization involves a certain ``hard thresholding.'' Another involves ``soft thresholding.''  We can estimate $C$ by the methodology of Chao (1987).  Denote her estimator by $\hat{C}$.  Then, given our list of approximations, the ``hard thresholding'' estimator of the common conditional covariance matrix of $\{ \BX_i\}$ is
$$[\hat{\theta}^* - (\hat{C}^{-2})] \BJ + [{\rm diag}\ (\BX' \BX) - (\hat{\theta}^* - (\hat{C}^{-2} ) \BI]_+ , $$
where $I$ is the $n\times n$ identity matrix.  For a square matrix $\BA$, diag ($\BA$) is the matrix with diagonal entries those of $\BA$, but with entries 0 elsewhere; for a real number $x$, max$(0, x) = x_+$.  The ``soft thresholding'' estimator of the common conditional covariance is
$$[\hat{\theta}^* - (\hat{C}^{-2})] \BJ + \ | {\rm diag}\ (\BX' \BX) - (\hat{\theta}^* - (\hat{C}^{-2} ) \BI |.  $$
We denote by vertical bars the matrix with absolute values of each entry for each entry.  In what follows, where we have done hard thresholding we use the term ``positive expectation.''  The soft thresholding is denoted by ``spherical shells.''  We have also tried another estimator available with ``corpcor.''  That is, cor.shrink; for details see the documentation for ``corp.cor''.

\section{Combining Estimators\label{section combiningEstimators}}

Readers please note that from our basic model that connects the unknown $\bp$ and our observations $\{\bp_i \}$, assumptions that elaborate on the model, and computations regarding derived quantities that owe, typically, to jackknifing or empirical Bayes estimation, we are able to estimate all quantities we require from $\{\bp_i \}$.  In this section we combine five estimators to yield our final estimates of clonality.  We denote the first of the five by $\hat{\theta}^{(0)}$.  It is the estimator of clonality that devolves from not regularizing the estimator (13).  We denote this ``unregularized'' estimator of covariance by $\hat{\bSigma}^{(0)}$. The estimator of $\theta$ is not exactly unbiased.  However, it is close and is one of five estimators we combine.  Another estimator of $\theta$ is exactly unbiased; that is $\hat{\theta}^*$.  To define what we term $\hat{\theta}^{(1)}, \hat{\theta}^{(2)}$, and $\hat{\theta}^{(3)}$ requires further definitions.  Remember that for every $k$, an estimator of
$E (||  \bfvarepsilon_k ||^2_2 \ | \ \bp, \{ C_i \})$ can be computed from $\bp_i$.  Write
$\bar{\varepsilon} = n^{-1} \sum^n_{m=1} E (||  \bfvarepsilon_m ||^2_2 \ | \ \bp, \{ C_l \})$, and write $\varepsilon_*$ for the minimum summand.  Now define
\begin{eqnarray}
\BT^{(1)} &= & \theta \ \bar{\psi}\ \bar{\varepsilon}\ \BI  \\ \label{eq15}
\BT^{(2)} &= &{\rm diag}\ (\bSigma^{(0)})  \\ \label{eq16}
\BT^{(3)}_{(k,l), (k', l')}
          &= &l' \hspace{5.7cm} \mbox{if}\ k, l, k', l'  \mbox{distinct} \nonumber \\
          &= &\theta \ \bar{\psi} \ 
              E (||  \bfvarepsilon_k ||^2_2 + ||  \bfvarepsilon_l ||^2_2 \ | 
              \ \bp, \{ C_i \}) 
              \ \ \mbox{if}\ (k, l) = (k', 0')  \\ \label{eq17}
	  &= & \theta \ \bar{\psi} \ \varepsilon_* \ \ \mbox{else} \nonumber 
\end{eqnarray}

We compute clonality now for the three estimators that devolve from 
$(1/2)(\hat{\bSigma}^{(0)} + \BT^{(1)})$, 
$(1/2) (\hat{\bSigma}^{(0)} + \BT^{(2)}),$ and 
$(1/2)(\hat{\bSigma}^{(0)} + \BT^{(3)})$.  Along with the first two estimators of $\theta$ we have cited, we have now five estimators in all.  They have a $5 \times 5$ covariance matrix, which we can estimate via jackknifing, again with replicates the sampling units.  Then, using the estimated $5\times 5$ covariance matrix and (14), we combine our five estimators of $\theta$ into final estimators.  There is one final estimator for each regularization of $(\hat{\theta}^* - C^{(-2}) \BJ$ in the previous section.  The final estimator is computed and returned by the ``inter.clonality'' function of our {\it lymphclon} package.  That ``inter.clonality'' function is part of the ``mixture.clonality'' component of the return value.

\section{Simulations\label{section simulations}}

We realize that readers who have read thus far may consider our ``final'' estimator(s) of clonality, $\theta$, extremely complex.  We caution them that with $\hat{\theta}^*$, the bar is extremely high.  (See Parameswaran et al., 2013); we believe that with accuracy measured by mean squared error, at present $\hat{\theta}^*$ is the best known to immunologists. Because at a particular point in the previous sections we were unable to fill in mathematical details, there and subsequently we relied on heuristic arguments.  That being the case, the only path we see to convincing any who read our works that our complex estimator(s) are ``good'' is to perform simulations.  While many details will be given in another paper, we give the results of some simulations here.

To begin, we require definitions.  The random variable $Z$ has a Zipf distribution with parameter 
$0 \le r \le 1$, if for nonnegative integers $j, P(Z=j) \propto  j^{-r}$.  Write $Z\sim$ Zipf($r$). With our choices of $r, \sum^\infty_{j=0} P(Z = j)$ is infinite, so we limit the upper limit of $j$, that limit here being $C$.  

The number of replicates, $n$, is specified by the user.  In what we report here, $n=8$.  We somewhat arbitrarily take $C=200,000$, though in {\it lymphclon} that number can be changed.  In {\it lymphclon}, each $\{C_i\}$ is chosen according to Zipf($r$) for various values of $r$.  In what we report here, the number of cells in six of the eight replicates was 1,000; in the other two it was 10,000.  Once $\{C_i\}$ are determined, then the number of cells assigned to particular rearrangements is simulated via a ``Pareto-Poisson machine.''  By this we mean that for each replicate we pick a Pareto random number with location and shape parameters 1.  Denote them by $P_1, ... P_n$.  We then simulate the assignments of cells to rearrangements according to Poisson distributions, with respective mean values $P_1, ... P_n$.  

Unconditionally, the cells by replicates have identical distributions.  So, the statistical notion of sufficiency entails that we might as well take $n=1$.  However, what matters to practice is that conditional on  $\{C_i\}$ and the assignments of cells to rearrangements, they do not.  

The wisdom of our admittedly complex configurations is our attempt to circumvent bias associated with amplification preferences and ``PCR jackpots'' (see Boyd et al., 2009).  The algorithm to which our complex estimator(s) of clonality are compared is that implied by (6) alone.  (See Parameswaram et al, 2013.) We do not model ``read errors'' or any bias in the PCR amplification.  

The one simulation we report here demonstrates remarkable results for our ``final'' estimators of clonality.  Readers can draw such conclusions as they wish regarding the different regularizations of our {\bf Section 5}.  \\


\newpage
\setcounter{page}{17}

\section{Summary\label{section summary}}
The goal of this paper is to demonstrate that from data on frequencies $\bp$ of V(D)J rearrangements, our new approaches can be combined with existing methodologies to estimate functionals of $\bp$ accurately. We quantify accuracy by mean squared error.  A takeaway message is that if the functional is quadratic in the entries of $\bp$, as is {\it clonality}, then pairs of quadratic functions of the data can be employed successfully.  Bias is removed by application of the jackknife when necessary.  Variance is reduced by empirical Bayes arguments, which enable estimation of key parameters.  These empirical Bayes arguments bring to mind but are more complex than empirical Bayes arguments for the James-Stein estimator (see Efron, 2010, and Efron and Morris, 1973).  

These V(D)J rearrangements are identified by sequencing of replicates, each of which owes to PCR amplification of an initial sample.  It is in the nature of problems we study that within replicates, some rearrangements are seen only once, and these rare rearrangements vary by replicate.  On the other hand, some rearrangements are seen many times, perhaps too many times. The beauty of our approach is that it does well, no matter the distribution of frequencies.

\centerline
{\bf Acknowledgments}

We thank Professor Daphne Koller for helpful discussions and for teaching Y.L. 

Funding was supported in part by grant 5U19AI090019 from the National Institute of Allergy and Infectious Diseases, and also by grant 4R37EB002784 from the National Institute of Biomedical Imaging and Bioengineering.

\centerline
{\bf References}
\renewcommand{\baselinestretch}{2.0}
\setlength{\parindent}{0cm}
\setlength{\parskip}{1em}

{Boyd, S.D., Marshall, E.L., Merker, J.D., Maniar, J.M., Zhang, L.N., Sahaf, B., Jones, C.D., Simen, B.B., Hanczaruk, B., Nguyen, K.D., Nadeau, K.C., Egholm, M., Miklos, D.B., Zehnder, J.L., and Fire, A.Z. ``Measurement and clinical monitoring of human lymphocyte clonality by massively parallel V-D-J pyrosequencing,'' {\it Science Translational Medicine} {\bf 1} 12a23 (2009). 

Breiman, L., Friedman, J.H., Olshen, R.A., and Stone, C.J. (1984) {\it Classification and Regression Trees}. Belmont, CA: Wadsworth International Group. 

Chao, A. ``Estimating the population size for capture-recapture data with unequal catchability,'' {\it Biometrics}  {\bf 43}(4)  (1987), 783--791.
 
Chao, A. ``Estimating population size for sparse data in capture-recapture experiments,'' {\it Biometrics} {\bf 45} (1989), 427--438.

Efron, B. (2010)  {\it Large Scale Inference Empirical Bayes Methods for Estimation, Testing, and Prediction.} Cambridge: Cambridge University Press. 

Efron, B. and Morris, C. ``Stein's estimation rule and its competitors—An empirical Bayes approach,'' {\it Journal of the American Statistical Association} {\bf 68} (1973), 117--130. 

Efron, B. and Thisted, R. ``Estimating the number of unseen species: How many words did Shakespeare know?'' {\it Biometrika} {\bf 63}(3) (1976), 435--447.

Hoeffding, W. ``Probability inequalities for sums of bounded random variables,'' {\it Journal of the American Statistical Association} {\bf 58} (1963), 13--30.  

Jiao, J., Venkat, K. and Weissman, T. ``Order-optimal estimation of functionals of discrete distributions,'' (2014) arXiv:1406.6956v2.

Jiao, J. Venkat, K. and Weissman, T. ``Maximum likelihood estimation of functionals of discrete distributions,'' (2014) arXiv:1406.6959v1.

Miller, R.G. ``The jackknife—A review,'' {\it Biometrika} {\bf 61}(1) (1974), 1--15.

Opgen-Rhein, R. and Strimmer, K. ``Accurate ranking of differentially expressed genes by a distribution-free shrinkage approach,'' {\it Statistical Applications in Genetics and Molecular Biology} {\bf 6}(1) (2007), 1--18.

Parameswaran, P., Liu, Y., Roskin, K.M., Jackson, K.K., Dixit, V.F., Lee, J.Y., Artiles, K.S., Zompi, S., Vargas, M.J., Simen, B.B., Hanczaruk, B., McGowan, K.R., Tariq, M.A., Pourmand, N., Koller, D., Balmaseda, A., Boyd, S.D., Harris, E., and Fire, A.Z. ``Convergent antibody signatures in human dengue,'' {\it Cell Host \& Microbe} {\bf 13}(6) (2013), 691--700.

Perlman, M.D. and Rasmussen, U. ``Some remarks on estimating a noncentrality parameter,'' {\it Communications in Statistics} {\bf 4} (1975), 455-468.

Qi, Q., Liu, Y., Cheng, Y., Glanville, J., Zhang, D., Lee, J-Y., Olshen, R.A., Weyand, C.M., Boyd, S., and Goronzy, J.J. ``Diversity and clonal selection in human T cell repertoire,'' (2014), submitted for publication.

Robbins, H.  ``An Empirical Bayes Approach to Statistics.''  {\it Proc. Third Berkeley Symp. on Math. Statist. and Prob.}  {\bf 1} (Univ. of Calif. Press, 1956), 157--163.

Robbins, H.E. ``Estimating the total probability of the unobserved outcomes of an experiment,'' {\it The Annals of Mathematical Statistics}'' {\bf 39}(1) (1968), 256--257. 

Sch$\ddot{a}$fer, J. and Strimmer, K. ``A shrinkage approach to large-scale covariance matrix estimation and implications for functional genomics.'' {\it Statistical Applications in Genetics and Molecular Biology} {\bf 4}(1) (2005), 1--30. 

Schatz, D.G. and Swanson, P.C. ``V(D)J recombination: Mechanisms of initiation,'' {\it Annual Review of Genetics} {\bf 45} (2011), 167--202.  

Varadhan, S.R.S., ``Large deviations,'' {\it The Annals of Probability} {\bf 36}(2) (2008), 397--419.

\end{document}